\documentclass[prb, preprint, aps,amsmath,showpacs, superscriptaddress]{revtex4}
\usepackage{graphicx}

\newcommand{\bv}[1]{{\bf #1}}

\begin{document}

\title{Monte Carlo simulations of ${\rm Ni Fe_2O_4 }$ Nanoparticles }

\author{Chenggang Zhou}
\affiliation{Computer Science and Mathematics Division, Oak Ridge National Laboratory, P. O. Box 2008, MS6493, Oak Ridge Tennessee, 37831-6493 USA } 
\affiliation{Center for simulational physics, University of Georgia, Athens Georgia, 30602 USA }

\author{T. C. Schulthess}
\affiliation{Computer Science and Mathematics Division, Oak Ridge National Laboratory, P. O. Box 2008, MS6493, Oak Ridge Tennessee, 37831-6493 USA }

\author{D. P. Landau}
\affiliation{Center for simulational physics, University of Georgia, Athens Georgia, 30602 USA } 
\date{\today}

\begin{abstract}
We use Monte Carlo simulations to study ${\rm Ni Fe_2O_4 }$  nanoparticles.
Finite size and surface effects differentiate them from 
their bulk counterparts. A continuous version of the Wang-Landau algorithm is used to calculate the joint 
density of states $g(M_z, E)$ efficiently. From $g(M_z, E)$, we obtain the Bragg-Williams 
free energy of the particle, and other physical quantities. 
The hysteresis is observed when the nanoparticles have both surface disorder and surface 
anisotropy. We found that the finite coercivity is the result of interplay between 
surface disorder and surface anisotropy. If the surface disorder is absent or the 
surface anisotropy is relatively weak, the nanoparticles often exhibit superparamagnetism.
\end{abstract}

\maketitle
\section{Introduction}
Many magnetic nanoparticles are found to be superparamagnets,\cite{Khanna91} since they
are not sufficiently big to contain more than one magnetic domain. 
On the other hand, a large fraction of the spins are located near the surface of the 
nanoparticle, and their local environments have lower symmetries than the bulk. 
These spins may prefer to align to directions different from the bulk ordering direction, 
thus reducing the total magnetization. This surface anisotropy might contribute 
significantly to the total magnetoanisotropy and the hysteresis of the nanoparticle,\cite{Frei57} 
e.g. ferrimagnetic ${\rm Ni Fe_2O_4 }$ nanoparticles. The bulk 
${\rm Ni Fe_2O_4 }$ is a ferrimagnet with a Neel temperature $T_N = 838$~K. 
The ball-milled nanoparticles have an average radius of $6.5$nm~\cite{Kodama99a} 
and have been previously found to show a slightly open hysteresis loop at low temperatures.\cite{Kodama96, Kodama97a}

In this paper, we use the heatbath algorithm and the Wang-Landau algorithm to calculate the 
finite-temperature properties of individual ${\rm Ni Fe_2O_4 }$ nanoparticles. 
The heatbath algorithm~\cite{Miyatake86, Loison04} runs at a certain temperature and external 
magnetic field. 
However, the heatbath algorithm samples does not directly 
generate the hysteresis, because the canonical ensemble it samples is dominated by  
the ground state rather than a meta-stable state, which contributes to the hysteresis. 
We use a continuous version of the Wang-Landau algorithm~\cite{Zhou05} 
to calculate the joint density of states of the nanoparticles, 
$g(M_z,E)$, where $M$ and $E$ are magnetization and 
internal energy of the nanoparticle respectively. $g(M_z,E)$ contains complete information of the
nanoparticle at any temperature and in any magnetic field. 

\section{Model of ${ \rm Ni Fe_2O_4 }$ nanoparticles and computational methods}

Bulk ${ \rm Ni Fe_2O_4 }$ has an inverted spinel structure of lattice constant 8.34~\AA. 
The eight tetrahedral sites in one unit cell are occupied by ${\rm Fe}^{3+}$, 
while eight ${\rm Fe}^{3+}$ and eight ${\rm Ni}^{2+}$ are distributed on the sixteen octahedral sites. 
We assume the ions on the octahedral sites have Verwey order, 
i.e. the ${\rm Fe}^{3+}$ and ${\rm Ni}^{2+}$ occupy (100) planes alternatively.  
In this system, $S =5/2$ for ${\rm Fe}^{3+}$ , and $S=1$ for ${\rm Ni}^{2+}$. The Hamiltonian
for the magnetic structure of the bulk consists of Heisenberg indirect exchange:
\begin{equation}
{\cal H}_{\text{bulk}} = -\sum_{<i,j>} J_{ij} \bv{S}_i \cdot \bv{S}_j - \bv{h} \cdot \sum_i \bv{S_i },
\label{ham}
\end{equation}
where the summation is over nearest and next nearest pairs with nonzero exchange $J_{ij}$. 
The exchange constants $J_{ij}$ are taken from Ref.~[\onlinecite{Kodama96}] and 
we use semiclassical approximation for the spins.  
In the ground state, ${\rm Fe}^{3+}$ spins on tetrahedral sites are antiparallel to ${\rm Fe}^{3+}$ 
spins on octahedral sites, while ${\rm Ni}^{2+}$ spins align with 
${\rm Fe}^{3+}$ spins on octahedral sites. The cubic anisotropy ($K_1 = -8.7\times10^4 {\rm erg/cm^3}$ 
at 77~K \cite{Ferrites}) is negligible for these nanoparticles. 

We model the nanoparticles by cutting off those spins outside a sphere of a given radius. 
The center of the sphere is randomly selected, which might not be a lattice point. 
Following the previous work\cite{Kodama96}, we randomly remove a small fraction of surface spins to 
model surface disorder, and use uniaxial anisotropy for the surface spins:
\begin{equation}
 {\cal H}_{\text{A}} = -k_s \sum (\hat{\bv{ s}}_i \cdot \hat{\bv{ n}}_i)^2,
\label{san}
\end{equation} 
where the unit vector $\hat{\bv{ s}}_i$ points in the direction of spin $\bv{S}_i$, 
and the unit vector $\hat{\bv{ n}}_i$ is chosen to be parallel to 
$ \sum_{j}^{nn}(\bv{r}_i - \bv{r}_j) $
where $\bv{r}_j$ are positions of nearest neighbors of $\bv{r}_i$ in the nanoparticle. 
As a result,
the surface anisotropy term vanishes if local inversion symmetry exists for that spin. 
$k_s$ in Eq.~(\ref{san}) characterizes the strength of the anisotropy. 
A realistic model would have $k_s$ dependent on the local environment of the spin. 
but in our simulations, $k_s$ is a positive tunable constant. 
We use 0, 2.5~K, 300~K, and 500~K for $k_s$, to observe the effect of different strength of 
the anisotropy.

We use the heatbath algorithm~\cite{Miyatake86,Loison04} and the improved Wang-Landau algorithm~\cite{Wang01, Zhou05} 
in our simulations.The heatbath algorithm reaches equilibrium in presence of disorder and anisotropy much faster than the Metropolis algorithm.\cite{Zhou04d} The continuous version of the Wang-Landau algorithm~\cite{Zhou05} calculates the joint density of states
\begin{equation}
  g(M_z,E) = \int \delta\left[{{\cal H}(\bv{S})\over N} - E \right] \delta\left( {\sum_i \bv{S}^z_i \over N} - M_z \right) 
  d \mu (\bv{S}), 
\label{rme}
\end{equation}
where $N$ is the number of spins, $\bv{S}$ collectively denotes all the spins in the model, 
$d \mu(\bv{S}) = \prod_i S_i^{-2}\delta(S_i^2 - \bv{S}_i^2) dS_i^x dS_i^y dS_i^z $ is the volume measure
of the phase space, $E$ and $M_z$ are the average energy per spin and average magnetization per
spin.
With $g(M_z,E)$ in Eq.~(\ref{rme}) at hand, one calculate the partition 
function, and the Bragg-Williams free energy defined as a function of temperature and magnetization:
$F(M_z,T) = - k_B T \ln \int g(M_z,E)e^{-\beta E N} dE$.
If the system exhibits a first order phase transition, $F(M_z,T)$ has double minima at constant $T$. The free energy
difference between the ground state and the metastable state as well as the free energy barrier between them are 
obtained from $F(M_z,T)$.

\section{Results and discussion}

\begin{figure}[b]
\includegraphics[width =0.9\columnwidth]{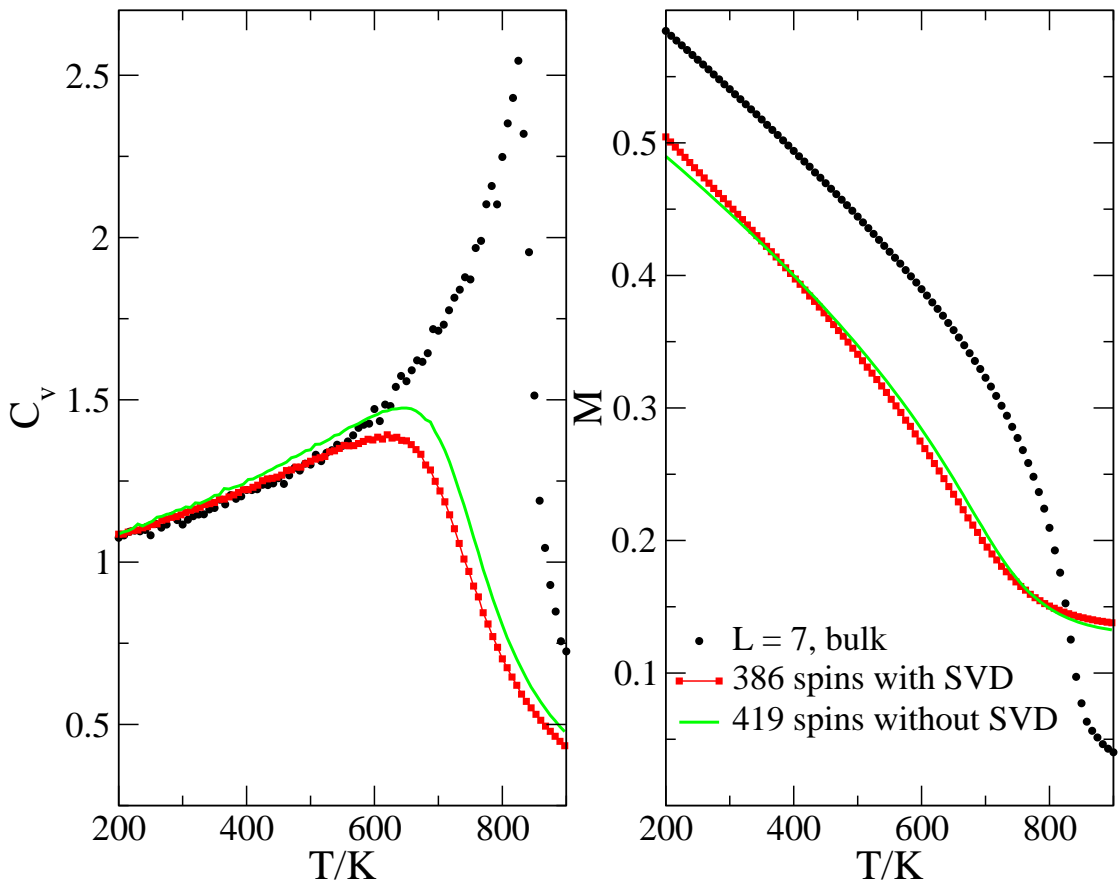}
\caption{ Comparison between the bulk material and the nanoparticle. Left: the
specific heat of nanoparticles has a smooth peak below the Neel temperature of
the bulk material. Right: the reduced magnetization of the nanoparticle. The bulk
sample is a cubic with 7 unit cells along each axis. }
\label{fig1}
\end{figure}
Using the heatbath algorithm, we calculated the specific heat and the Binder cumulant of 
magnetization for bulk samples of different sizes with periodic boundary condition. With finite
size scaling, we obtain the Neel temperature for ferrimagnetic phase $T_N = 869K$, which is
about 4\% higher than the experimental value. We also compared the simulations for bulk material 
to those for the nanoparticles. 
Figure~\ref{fig1} shows the effect of the size and surface of the nanoparticle. 

The peak in the specific heat of the nanoparticle 
appears at a lower temperature, about 200~K below that of the bulk material. 
The right panel shows the magnetization defined as
$ M = N^{-1} \left| \sum_{i=1}^N \bv{S}_i \right| $.
The magnetization of the nanoparticle is substantially suppressed. Near the 
transition temperature, the magnetization of the nanoparticle gradually ramps 
up with decreasing temperature, while the bulk materials show a magnetization 
curve typical of the ferrimagnetic phase transition.

\begin{figure}[b]
\includegraphics[width =0.9\columnwidth]{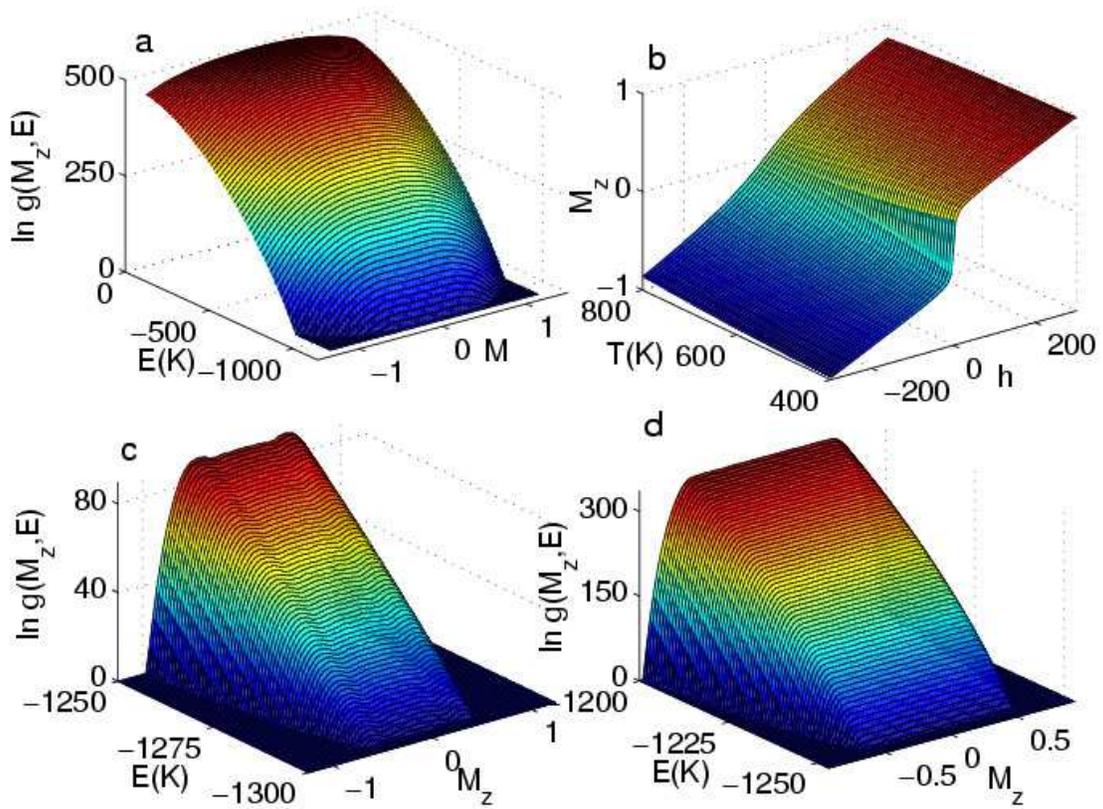}
\caption{Density of states and magnetization of a typical nanoparticle containing 387 spins. 
The normalization constant in the density of states is arbitrarily selected. 
See the text for detailed explanations.}
\label{fig2}
\end{figure}

Figure~\ref{fig2} shows $g(M_z,E)$ for a typical nanoparticle with surface disorder 
and 387 spins. Figure~\ref{fig2}(a) shows $g(M_z,E)$ for a large
energy range: $-1200$~K$<E<0$~K, which illustrates the generic behavior
of density of states of these nanoparticles. Here the magnetization $M_z$ is the average 
$z$-component of the spins.
At high energies, $g(M_z,E)$ has a single maximum value at $M_z=0$ for constant $E$. 
At low energies, this single maximum is replaced by a relatively flat part, 
i.e. $\partial g(M_z,E) /\partial M_z \approx 0$, which indicates a non-zero
spontaneous magnetization. 
The energy range of Fig.~\ref{fig2}(a) is sufficient for calculating temperature
dependent properties for approximately $T>400$~K.
The magnetization as a function of temperature and external field is shown in 
Fig.~\ref{fig2}(b). A spontaneous magnetization develops at about $T=700$~K, 
consistent with the reduced transition temperature shown in Fig.~\ref{fig1}.  
Since the model has continuous degrees of freedom, $g(M_z,E)$ is logarithmically 
divergent near the ground state energy.\cite{Zhou05} 
Thus, the lowest energy that a simulation can efficiently sample
is restricted by its resolution in energy. Fig.~\ref{fig2}(c) and (d) show
$g(M_z,E)$ at low energies  calculated with higher resolution in energy.
Their energy ranges allow a temperature range $180$~K$<T<220$~K when 
the $g(M_z,E)$ is used to calculate temperature dependent quantities.
In Fig.~\ref{fig2}(c),  a double-valley structure appears between
$M = -0.4$ and $M=0.4$. 
\begin{figure}[t]
\includegraphics[width =0.9\columnwidth,height=0.8\columnwidth]{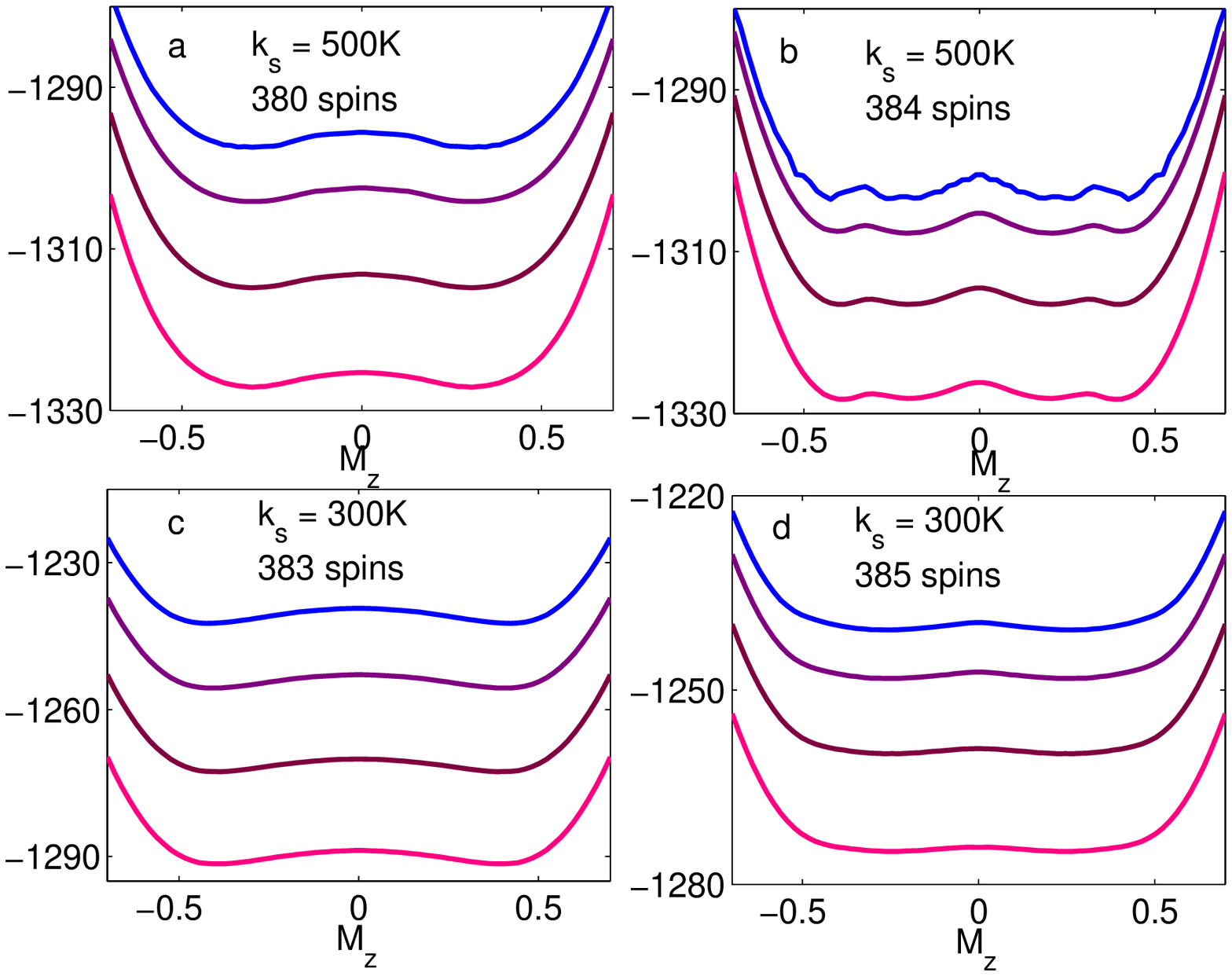}
\caption{Bragg-Williams free energy per spin, a variety of structures are observed in 
nanoparticles with strong surface anisotropy and disorder, at relatively high temperature. 
In each panel, four free energy curves are for temperatures 187~K, 
216~K, 244~K, and 272~K respectively from top to bottom.}
\label{fig3}
\end{figure}
This structure is responsible for a finite coercivity,
because the magnetization is blocked by it  when the magnetic field is slightly reversed. 
The double valley structure becomes more pronounced at lower temperatures.
On the contrary, in Fig.~\ref{fig2}(d), the density of states
of the same nanoparticle with a reduced surface anisotropy constant $k_s = 2.5$~K, 
does not show any noticeable structure as in Fig.~\ref{fig2}(c). 
The energy range in (d) is higher than that in (c) because of an over all energy shift caused by the 
change in surface anisotropy. However the energy range of (d) corresponds to a 
temperature range of $40$~K$<T<80$~K.

The lack of a finite coercivity in Fig.~\ref{fig3}(d) is due to the 
result of the rotational symmetry. 
Obviously the surface disorder preserves the rotational symmetry. 
Although there is a small $k_s$, the surface anisotropy is too weak to destroy the rotational symmetry
 at temperatures much larger than $k_s$. 
Consequently, the nanoparticle is well described by superparamagnetism.

When we assume a strong surface anisotropy, which is ``larger'' than the thermal energy 
$kT$ and comparable to the exchange field on the surface, then we indeed observe various 
structures in $g(M_z,E)$,  which are responsible for coercivity at finite temperatures. 
Figure~\ref{fig3} shows four examples of this 
behavior. We plot the Bragg-Williams free energy per spin $F(M_z,T)/N$ in Fig.~\ref{fig3}. 
Clearly, all of these four examples appear to have a barrier of free energy between two states 
of spontaneous magnetization, also small oscillations on top of the barrier is
shown in Fig.~\ref{fig3}(b). 

To observe the effect of the intermediate surface anisotropy, 
we used $k_s = 50$~K and performed the 
calculation with the same set of nanoparticles that we used for the weak and strong
surface anisotropy. We found most of them do not appear to have a noticeable depression
in the density of states or a peak the free energy even at low temperatures with a few 
exceptions. One of them is the nanoparticle used to do the calculation for Fig.~\ref{fig3}(d), 
it was found that with $k_s = 50$~K, its Bragg-Williams free energy for 
$35$~K$<T<55$~K appears to have the similar shape as shown in Fig~\ref{fig3}(d).

In summary, we have studied the magnetic properties of ${\rm Ni Fe_2O_4 }$ 
nanoparticles using two  kinds of Monte Carlo simulations.
We have found that only both surface disorder and a pretty strong
surface anisotropy are required to develop the hysteresis. 
We have also noticed that a strong surface anisotropy is 
required because our model has a near-spherical shape. 
The surface anisotropies at different positions are likely to cancel each 
other leaving a very small overall anisotropy. 
In fact, we have observed that the total magneto-anisotropy energy is only of 
order $0.1k_s$. Our approach is purely based on thermodynamics and presents no information 
on the spin dynamics of the nanoparticle. It is very likely that the surface spins form a 
disordered glass-like layer, which slows down the reversal of magnetization.

\section{Acknowledgements}

This research is supported by the Department of Energy through the Laboratory Technology 
Research Program of OASCR and the Computational Materials Science Network of BES under 
Contract No. DE-AC05-00OR22725 with UT-Battelle LLC, and also by NSF DMR-0341874.


\end{document}